\documentclass[conference]{IEEEtran}
\IEEEoverridecommandlockouts
\usepackage{amsmath,amssymb,amsfonts}
\usepackage{algorithmic}
\usepackage{textcomp}
\usepackage{xcolor}
\usepackage{stfloats}
\usepackage{comment}
\usepackage{subcaption}
\usepackage{threeparttable}
\usepackage{xfrac}
\usepackage{comment}
\usepackage{hyperref}
\usepackage{tabto}
\usepackage{orcidlink}
\usepackage{textcomp}
\usepackage{cite}

\begin{document}

\bstctlcite{IEEEexample:BSTcontrol}

\newcommand{\ipv}[1]{{\color{blue}[IPV: #1]}}

\title{Dynamic Power Management Methodology for Distributed Vertical Power Delivery in High-Performance Computing Systems\\
%{\footnotesize \textsuperscript{*}Note: Sub-titles are not captured in Xplore and should not be used}
\thanks{This work was supported by the Center for Heterogeneous Integration of Micro Electronic Systems (CHIMES), one of seven centers in Joint University Microelectronics Program (JUMP) 2.0, a Semiconductor Research Corporation (SRC) program sponsored by the Defense Advance Research Project Agency (DARPA).}}

\author{
\IEEEauthorblockN{Sriharini Krishnakumar}
\IEEEauthorblockA{
\textit{University of Illinois Chicago}\\
Chicago, IL, USA \\
skrish47@uic.edu}
\and
\IEEEauthorblockN{Inna Partin-Vaisband}
\IEEEauthorblockA{ 
\textit{University of Illinois Chicago}\\
Chicago, IL, USA \\
vaisband@uic.edu}
}

\maketitle

\begin{abstract}

Distributed vertical power delivery (DVPD) architectures employ multiple parallel voltage regulators (VRs) to meet the high-power and high-current-density demands of modern high-performance computing (HPC) systems. While full parallel activation maximizes efficiency near peak load, medium–light load operation leads to efficiency degradation when all VRs remain active due to persistent switching and gate-drive losses. This work proposes a load-aware power system activation framework targeted at the medium–light load regime, in which the number of active VRs scales proportionally with instantaneous load power. A spatially informed selection strategy determines which VRs are activated from the available pool, aligning regulator placement with localized power demand. This locality-aware activation minimizes lateral redistribution currents within the power plane and reduces conduction losses and voltage drops. Simulation results on a representative DVPD system demonstrate 2\textendash3$\times$ switching-loss reduction relative to conventional full-parallel light-load operation, while sustaining an approximately 87\% efficiency plateau across the 5\textendash30\% load range. Output ripple constraints are preserved, with inductor current ripple maintained within 6\% and output voltage ripple within 2\%, ensuring regulation integrity while improving overall conversion efficiency.

\end{abstract}


\section{Introduction}

Power demands in modern high-performance computing (HPC) systems continue to increase, with individual chips exceeding 2~kW and wafer-scale platforms approaching 50~kW \cite{khairy2020,HIR2024}. Correspondingly, average current densities have escalated to 2–4~A/mm\textsuperscript{2}, requiring power delivery architectures capable of maintaining efficiency above 90\% across a wide operating range, typically 5–100\% of thermal design power (TDP) \cite{radhakrishnan2021power}. Efficient power delivery in these systems is enabled by distributed vertical power delivery (DVPD) architectures, which employ laterally distributed voltage regulators (VRs)  with vertically integrated power switches, inductors, and capacitors \cite{krishnakumar2023vertical,krishnakumar2024design,khorasani2025embedded,rasheedi2025high,rasheedi2024embedded}.

\begin{figure}[t!]
\vspace{-10pt}
\centerline{\includegraphics[width=0.5\textwidth]{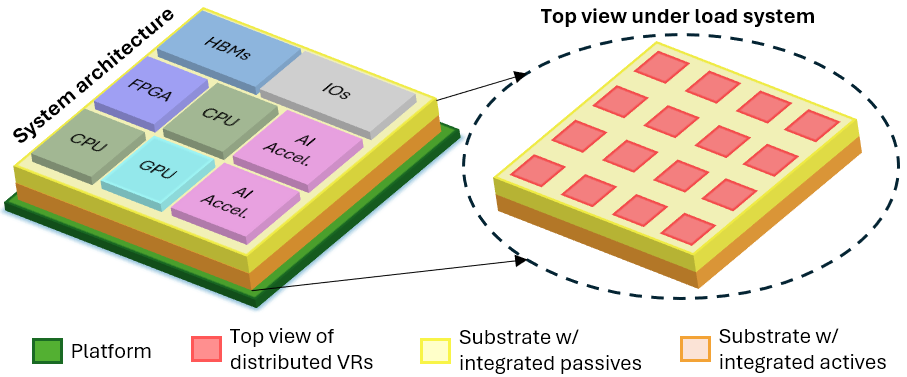}}
\caption{Schematic of a DVPD system.}
\label{fig:DVPD}
\vspace{-15pt}
\end{figure}

In a DVPD system, as illustrated in Fig.~\ref{fig:DVPD}, multiple VRs are connected in parallel to a common low-voltage power plane that supplies spatially distributed heterogeneous loads such as graphics-processing-units (GPUs), control-precessing-units (CPUs), high-bandwidth memory (HBMs), accelerators and other loads, mounted on a shared substrate. Each VR delivers a fraction of the total load current, enabling current sharing across multiple conduction paths. This architecture reduces electrical distance to the points of load (POLs), lowers parasitic impedance, and improves transient response while supporting high power-density operation.

However, conventional DVPD designs primarily optimize conversion and power delivery network efficiency for the maximum rated power $P_{\text{max}}$, where thermal constraints dominate the design space \cite{abdelzaher2025hybrid,khorasani2024package, krishnakumar2024vertical,krishnakumar2024system}. In a system comprising $M$ distributed VRs, each regulator is typically designed to deliver $P_{\text{max}}/M$. At this operating point, conversion efficiencies above 85\% are achievable while maintaining thermal hotspots below 85~°C \cite{choi2024thermal,choi2025substrate,choi2025self,choi2025automated}.

\begin{figure*}[t!]
\centerline{\includegraphics[width=\textwidth]{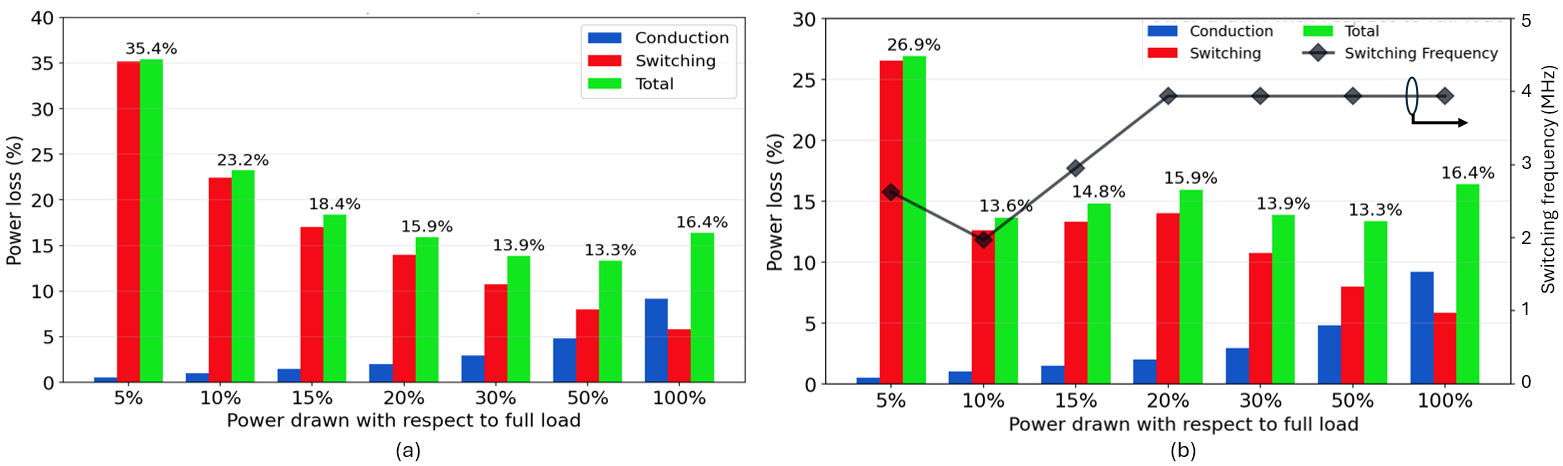}}
\caption{Typical DVPD loss as a function of percent load power for (a) fixed-frequency PWM, and (b) pulse-frequency modulation (PFM) operation. Loss components are expressed as a percentage of the total input power. Conduction losses include supply-to-VR routing, VR-to-POL routing, internal VR routing, inductor conduction, and power-switch conduction losses.}
\label{fig: PWM_PFM}
\vspace{-15pt}
\end{figure*}

In practical operation, however, power demand is highly dynamic and workload dependent. Utilization frequently deviates from peak load due to scheduling policies, workload variability, and maintenance intervals, leaving compute nodes idle for extended periods \cite{10.1145/3291606,ilsche2024optimizing}. Although techniques such as dynamic voltage and frequency scaling (DVFS) and power gating reduce compute-domain idle power \cite{abdelzaher2025bandit}, inefficiencies within the power delivery architecture remain largely unaddressed. Under such light-load conditions with DVPD systems, the conduction losses decrease with current, whereas switching-related losses persist under fixed-frequency operation, resulting in degraded efficiency and reduced power quality\cite{trescases2011survey,fei2016two}. For example, delivering 25\% TDP using four VRs designed for 100\% TDP operation results in each module operating far from its optimal efficiency point, while fixed-frequency switching losses remain largely unchanged. In contrast, consolidating the same load into a single VR operating near its peak-efficiency region (and turning off the remaining VRs) yields higher overall efficiency. For large-scale HPC systems operating below peak load for a significant fraction of runtime, even modest efficiency degradation translates to substantial system-level energy overhead.

Despite advances in converter-level light-load control, current DVPD architectures lack a system-level mechanism to dynamically include or exclude regulators in response to instantaneous load magnitude and spatial distribution. Selective regulator deactivation in a distributed architecture must preserve system-wide efficiency, regulation, and power quality while avoiding excessive redistribution currents within the power plane. To address this limitation, this work introduces a load-aware power system activation framework for DVPD architectures. The proposed approach dynamically scales the number of active VRs according to instantaneous load power and spatial demand distribution, maintaining near-optimal per-regulator operation while mitigating unnecessary switching losses.
The main contributions of this work are:

\begin{itemize}
    \item A spatially coordinated power-management scheme for DVPD systems that differs fundamentally from conventional phase shedding by selectively activating distributed VR modules to preserve current locality and minimize redistribution currents within power distribution network (PDN).

    \item A load-aware activation strategy that proportionally scales the number of active VRs with instantaneous load power while maintaining ripple and regulation constraints. %An architectural load-aware activation framework that proportionally scales active VRs with load power in DVPD systems.
    
    \item Demonstration of substantial switching-loss reduction and near-flat efficiency scaling in the medium–light load regime.
\end{itemize}

%To sustain high conversion efficiency while preserving power quality under light-load operation, this work proposes a load-aware power system activation methodology for DVPD systems. The proposed framework dynamically adjusts the number of active VRs based on spatially and temporally varying load conditions, redistributing current to maintain near-optimal per-VR operation.
%To evaluate the proposed methodology, a representative DVPD architecture is analyzed using a coupled SPICE–Python co-simulation platform. Detailed electrical models of the VRs, load segments, and power delivery network are implemented in SPICE, while a Python-based optimization engine monitors spatially and temporally varying load conditions to determine optimal VR activation states in real time.

The remainder of this paper is organized as follows. Conventional fixed-frequency pulse-width-modulation and pulse-frequency-modulation operation in DVPD systems are reviewed in Section~II. Their impact on light-load efficiency and power quality is also evaluated in Section~II. The proposed load-aware activation methodology is presented in Section~III, along with simulation results demonstrating efficiency, ripple current, and output voltage ripple across operating conditions. The paper is concluded in Section~IV.

\section{Existing Control Strategies} \label{sec:control}

A representative DVPD system comprising 70 integrated VRs delivering 1~kW at 1~V is used throughout this work to evaluate various power control strategies. Each VR performs a 48~V $\to$ 1~V conversion, delivers up to 15~A, and operates at a nominal switching frequency of 4~MHz. The peak-to-peak inductor current ripple is designed to be 6\% of rated current ($\Delta i = 0.825$~A), with a regulated output voltage ripple of $\Delta v = 0.02$~V.

\subsection{Pulse Width Modulation (PWM)}

The power-loss distribution for the representative system under fixed-frequency PWM is shown in Fig.~\ref{fig: PWM_PFM}a. Peak efficiency (86\%) occurs near 50\% load, where conduction and switching losses are balanced. %Efficiency degrades both above and below this operating point—at higher load due to increasing conduction losses, and at reduced load due to the persistence of switching and gate-drive losses. 
At higher loads, efficiency degrades due to increasing conduction losses. At reduced loads, conduction losses decrease with current, but fixed-frequency switching and gate-drive losses remain approximately constant, causing them to dominate total dissipation and reduce overall efficiency.
Thus, efficiency of the system degrades above and below the 50\% operating point. The degradation at light load is particularly pronounced, with the overall efficiency dropping to 77\% at 10\% of full-load, despite the substantially reduced load power. 

As load current decreases, the absolute current ripple $\Delta i$ remains approximately constant, whereas the relative ripple ${\Delta i / I_{\text{load}}}$ increases to approximately 60\%, degrading power quality. Although magnitude of output voltage ripple remains regulated, the output capacitor must sustain higher ripple current, resulting in additional self-heating and reduced component lifetime. 

This behavior indicates that under light-load operation, a significant fraction of input power is dissipated in switching mechanisms rather than delivered to the load. Consequently, fixed-frequency PWM is inherently inefficient at reduced load.

%This baseline behavior indicates that a significant fraction of input power under light-load conditions is dissipated in switching-related mechanisms rather than delivered to the load. Consequently, conventional fixed-frequency PWM control is inherently inefficient at reduced load. To alleviate the switching-loss dominance under light-load conditions, alternative control strategies have been proposed. These approaches can be broadly classified into pulse frequency modulation, which reduce switching activity and often permit discontinuous-conduction-mode (DCM) operation, and continuos-conduction-mode (CCM) based techniques, which preserve conduction characteristics by adjusting switching frequency. Each class offers distinct trade-offs between efficiency and power quality. For the representative DVPD system shown in Fig. \ref{fig: Loss distribution} , peak
%efficiency occurs near 50\% load, whereas efficiency falls below
%83\% at approximately 25\% of full load. Imposing an 83%
%minimum efficiency constraint therefore defines the transition
%from fixed-frequency PWM to PFM at this operating point.

\subsection{Pulse Frequency Modulation (PFM)}

In PFM, the switching frequency $f_{\text{sw}}$ is decreased with load current to reduce switching activity, thereby thereby reducing switching losses under light-load conditions. 
For the representative DVPD configuration with PWM, efficiency decreases below approximately 84\% at $\sim$20\% load.
%For the representative DVPD system, efficiency falls below 83\%  when approximately 25\% of full-load power is drwan. 
Imposing an 84\% minimum efficiency constraint therefore defines the PWM-to-PFM transition point. Note that, 84\% is selected as a design reference threshold to motivate transition to alternative control strategies while preserving ripple and regulation constraints; it does not represent a universal limit but rather a system-level design choice for the specified architecture and operating conditions.

While in continuous-conduction-mode (CCM), lower $f_{\text{sw}}$ improves efficiency but increases ripple magnitudes. The inductor current ripple scales linearly as ${\Delta i \propto 1/f_{\text{sw}}}$, while the output voltage ripple scales quadratically as  ${\Delta v \propto \Delta i/f_{\text{sw}} \propto 1/f_{\text{sw}}^2}$. Consequently, voltage ripple increases rapidly with PFM, degrading power quality.

As the ripple amplitude approaches the average inductor current, the VR transitions from CCM into boundary conduction mode (BCM), and subsequently into discontinuous-conduction-mode (DCM). In DCM, the voltage conversion ratio of a buck converter is given by

\begin{equation}
\frac{V_o}{V_{in}} =
\frac{2}{1+\sqrt{1+\frac{4K}{D^2}}},
\end{equation}
where $D = T_{on}/T_s$ is the duty cycle, $T_s = 1/f_{\text{sw}}$ is the switching period, ${K = 2L / R_L T_s}$, $L$ is the inductance, and $R_L$ is the load resistance. For regulated output voltage, the term $\frac{4K}{D^2}$ must remain approximately constant. As load current decreases, voltage regulation can be achieved by either reducing the on-time $T_{on}$ or the switching frequency.

Reducing $T_{on}$ %is feasible in low conversion-ratio designs where sufficient duty-cycle margin exists. However, 
in high conversion-ratio systems (e.g., 48~V to 1~V) is however impractical since the nominal duty cycle is inherently small (e.g., ${D \sim V_o/V_{in} \sim 1/48}$), resulting in extremely short on-times under multi-megahertz operation. Practical driver and device constraints impose a minimum achievable $T_{on}$, leaving negligible margin for further reduction at light load. Consequently, regulation under DCM is primarily achieved by decreasing switching frequency.

Fixed-on-time PFM implements frequency reduction by extending the off-time interval, effectively skipping switching pulses \cite{ilsche2024optimizing}. While this improves light-load efficiency, output voltage ripple is increased due to enlarged inductor current excursions. The load-dependent switching frequency introduces spectral spreading, complicating electromagnetic interference (EMI) mitigation. Furthermore, the achievable transient response is limited, since fewer switching events are available to correct load disturbances.

In distributed VR architectures supplying tightly regulated sub-1~V loads, DCM operation is incompatible with stringent ripple, transient, and EMI constraints. Therefore, frequency scaling should be bounded to CCM. 
%The switching frequency is reduced only to the minimum value that satisfies the CCM boundary for a given load, preventing entry into DCM while providing partial efficiency improvement. 
Such bounded frequency behavior is illustrated in Fig.~\ref{fig: PWM_PFM}b. As the load decreases below the PWM-to-PFM transition point (approximately 20\% of full load in the representative DVPD system), the switching frequency is reduced proportionally to maintain efficiency above the imposed 83\% constraint. Frequency scaling continues with decreasing load until the CCM boundary is reached at approximately 10\% load. Beyond this point, further reduction in $f_{\text{sw}}$ would force DCM operation; therefore, the switching frequency is maintained at or above the minimum value required to preserve CCM.

Consequently, efficiency improvement saturates at light loads because switching losses can no longer be reduced without violating ripple and transient constraints. Although bounded PFM provides partial recovery of light-load efficiency, the output voltage ripple increases to approximately 0.1~V (10\% of the 1~V output) at light-load, which exceeds typical ripple specifications for tightly regulated sub-1~V loads and imposes additional stress on downstream loads.

%\begin{figure}[t!]
%\centerline{\includegraphics[width=0.5\textwidth]%{figures/PFM.png}}
%\caption{Schematic of a DVPD system.}
%\label{fig: PFM}
%\vspace{-15pt}
%\end{figure}

%In fixed-on-time PFM control, regulation is achieved by modulating the switching frequency such that the average delivered power matches the load demand. With a fixed $T_{on}$, frequency reduction is accomplished by extending the off-time interval, effectively skipping switching pulses. This pulse-skipping improves light-load efficiency, but the operation in DCM increases output voltage ripple due to the enlarged inductor current excursions between switching events. The load-dependent switching frequency further introduces spectral spreading, which may complicate electromagnetic interference (EMI) mitigation. Moreover, operation at reduced frequency limits the achievable transient response, since fewer switching events are available to correct load disturbances. These effects restrict the applicability of fixed-on-time PFM in high-power DVPD systems for HPC platforms, where tight voltage regulation, fast transient response, and predictable EMI behavior are required.

%PFM operation within CCM remains acceptable. However, DCM operation is undesirable in distributed VR architectures. Therefore, frequency scaling must be bounded such that the converter remains in CCM. The switching frequency is reduced only to the minimum value that satisfies the CCM boundary for a given load, preventing entry into DCM while still providing partial light-load efficiency improvement.

\section{Load Aware Power System Activation (LAPSA)}

\begin{figure*}[t!]
\centerline{\includegraphics[width=\textwidth]{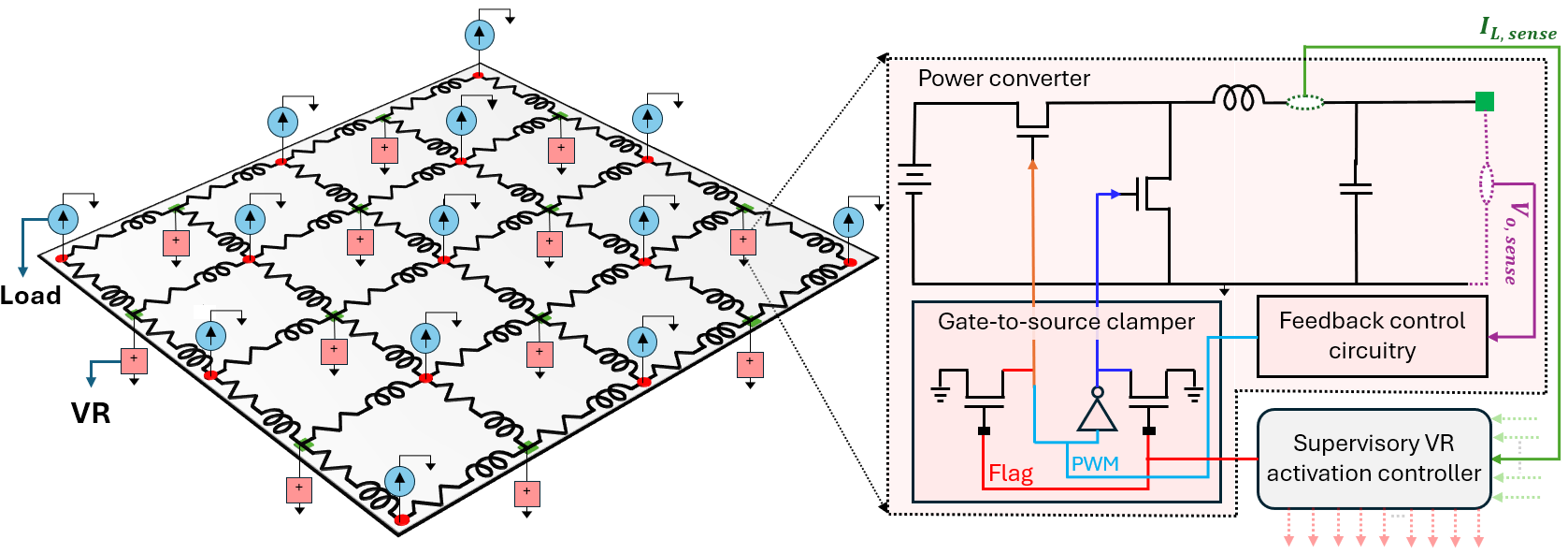}}
\caption{Load-aware power system activation framework for DVPD systems. The left panel illustrates the spatially distributed VR–load network across the power plane, where regulators supply localized load segments. The right panel shows the regulator-level implementation, including the gate-to-source clamp network and supervisory control interface.}
\label{fig: Framework}
\vspace{-10pt}
\end{figure*}

Unlike centralized power delivery architectures, DVPD supplies the load through multiple parallel VRs connected to a common low-voltage power plane. Under peak-load conditions, all regulators are required to satisfy current-density and thermal constraints. However, at light load, this parallel architecture introduces spatial redundancy. When load current decreases, maintaining all VRs in CCM degrades efficiency, since each regulator operates far below its optimized power level. In this regime, per-VR switching losses remain approximately constant, while delivered output power decreases, thereby reducing conversion efficiency.

Supplying reduced load demand using only a subset of regulators operating near their peak-efficiency workload points, while temporarily deactivating the remaining units, is proposed in this work. This selective activation improves system-level efficiency without violating peak current-density or thermal constraints. Based on this principle, a \textit{Load-Aware Power System Activation (LAPSA)} framework is introduced as a system-level power-management methodology for DVPD architectures. Unlike conventional phase shedding, LAPSA dynamically coordinates both the number and spatial distribution of active VR modules in response to workload variations across space and time. High power-density regions are supported by activating additional nearby VRs, while lightly loaded regions operate with fewer active units. By concentrating current within the active subset, each regulator operates closer to its optimal efficiency region, improving the balance between conduction and switching losses and enhancing overall conversion efficiency.

Unlike frequency-modulation techniques that alter switching frequency and disturb ripple characteristics, nominal VR load conditions are preserved with LAPSA. Each active VR continues operating at the designed switching frequency while maintaining the intended ripple envelope. Consequently, output ripple magnitude, control-loop stability, and EMI characteristics remain unchanged.

%In contrast to PFM-based approaches, which reduce switching activity within every regulator, LAPSA reduces the number of actively switching regulators. Therefore, switching and gate-drive losses are incurred only by the enabled regulators, causing the aggregate switching overhead to scale with the active subset rather than the total deployed VR count.

\subsection{Activation Rule and Control Implementation}

Let $P(t)$ denote the instantaneous load power and let $P_{\text{opt}}$ represent the load power corresponding to peak system efficiency under nominal operation with all $M$ regulators active. The activation policy maintains each enabled regulator near its optimal operating region by proportionally scaling the number of active units with load power. The required number of active regulators is expressed as

\begin{equation}
N_{\text{act}}(t)=
\begin{cases}
M, & P(t)\ge P_{\text{opt}},\\[6pt]
\left\lceil M \dfrac{P(t)}{P_{\text{opt}}}\right\rceil, & P(t)<P_{\text{opt}}.
\end{cases}
\label{eq:Nact_piecewise}
\end{equation}

This proportional activation maintains the per-regulator load near $P_{\text{opt}}/M$, maintaining near-optimal efficiency, regulated voltage, and ripple constraints across a broad range of light loads.

At each time, $N_{\text{act}}(t)$ regulators are enabled from the total pool of $M$ available VRs, while the remaining ${M - N_{\text{act}}(t)}$ VRs are disabled. The selection of active VRs is spatially informed: regulators located near regions of elevated load are preferentially enabled, thereby reducing lateral redistribution currents within the power plane. This proximity-aware allocation mitigates power plane conduction losses, steady-state IR drop, and transient inductive voltage excursions under spatially non-uniform power load.

Binary activation (i.e., full enable or complete disable of a VR module) is implemented using a gate-to-source clamp network controlled by a deactivation flag (see Fig. \ref{fig: Framework}). The gates of both the high-side and low-side switches of power converters are tied to the drain of an auxiliary NMOS transistor whose source is grounded. When the deactivation flag is asserted, the auxiliary NMOS pulls the gate voltage to ground, enforcing $V_{GS}=0$~V and overriding the PWM drive signals to fully disable switching activity. When the flag is deasserted, the clamp is released and the switches resume nominal fixed-frequency operation with $V_{GS}=5$~V.

\subsection{Co-Simulation Framework}

The proposed framework is evaluated using a coupled SPICE--Python co-simulation platform as described in Fig. \ref{fig: Framework}, that integrates device-level electrical modeling with supervisory control execution. Each VR, load segment, and interconnect path is modeled in SPICE using behavioral representations of power switches, passive elements, and parasitic interconnect impedances. Spatially and temporally varying HPC workload conditions are emulated to reproduce realistic load transients across the DVPD network.

Embedded current-sensor models are used to estimate the instantaneous output current of each VR. Practical on-chip current-sensing circuits and digital control hardware for such applications are well established and are not the primary focus of this work \cite{vaisband2015power}. The sensed data is exported to a Python-based supervisory VR activation controller, which computes $N_{\text{act}}(t)$ according to (\ref{eq:Nact_piecewise}) and determines the corresponding (de)activation flags for each converter in the DVPD system. These flags are then communicated back to the SPICE environment to reconfigure the active regulator subset during simulation.

%Reconfiguration latency is governed by three factors: (i) the execution time of the Python supervisory algorithm, (ii) communication overhead between the SPICE and Python environments, and (iii) the intrinsic response time of the gate-level control circuitry. In the present implementation, the Python execution requires approximately 2.7~s of wall-clock simulation time, while the data exchange between SPICE and Python introduces an additional 2~s. 

Reconfiguration latency in a practical implementation is governed by three components: (i) load sensing and digitization latency, (ii) execution time of the embedded supervisory VR activation controller, and (iii) gate-level propagation delay associated with enabling or disabling individual VRs. Load sampling is typically performed at microsecond or sub-microsecond intervals depending on ADC resolution and controller bandwidth \cite{wang2007design}. The activation decision logic can be executed within tens to hundreds of nanoseconds in FPGA- or ASIC-based digital control hardware \cite{halivni2020digital}. Communication latency between the supervisory controller and local VR enable registers is expected to be on the order of tens of nanoseconds in a centralized bus architecture \cite{kondoro2021low}. Once the updated activation state is issued, the hardware-level reconfiguration—implemented through the gate-to-source clamp network—occurs within 10~ns, corresponding to the simulated propagation delay, of the auxiliary control transistor. It is important to note that the supervisory layer modifies only the activation state of each VR, it does not alter duty ratio, switching frequency, or compensation parameters. Leakage losses of disabled VRs are included in the power loss model and are negligible relative to the switching-loss reduction achieved through the LAPSA selective activation.

\subsection{Efficiency Evaluation}

\begin{figure}[t!]
\centerline{\includegraphics[width=0.5\textwidth]{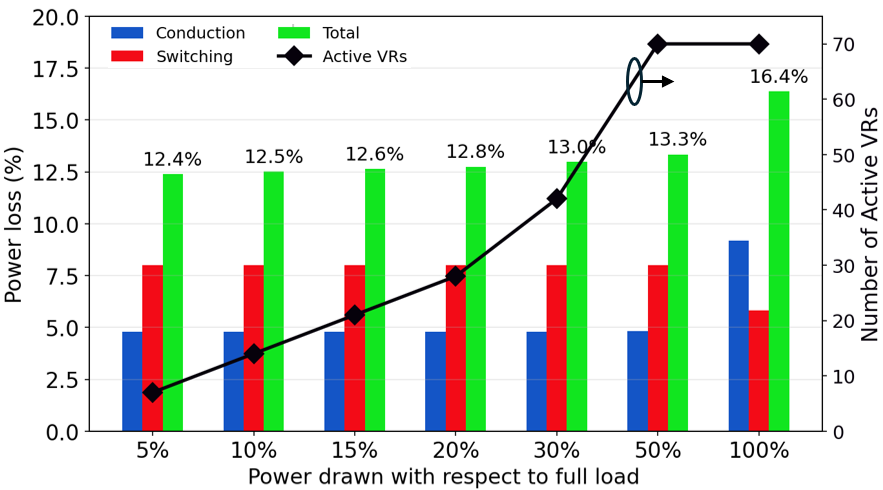}}
\caption{Loss distribution and active VR count under the proposed LAPSA framework.}
\label{fig: LAPSA}
\end{figure}

The efficiency improvement achieved with the proposed LAPSA methodology is demonstrated for the representative DVPD system described in Section~\ref{sec:control}, as shown in Fig.~\ref{fig: LAPSA}. Under baseline fixed-frequency PWM operation without light-load optimization, peak efficiency occurs at 500~W (50\% load), as illustrated in Fig.~\ref{fig: PWM_PFM}. This power load defines the LAPSA activation threshold.

When the consumed power decreases below 500~W, the number of active VRs is proportionally reduced according to (\ref{eq:Nact_piecewise}), resulting in seven active regulators at 5\% load. This proportional deactivation mitigates light-load efficiency degradation and maintains system-wide efficiency above 85\% across the 5--50\% load range, with less than 0.6\% variation. These results demonstrate near-flat efficiency scaling across light-to-medium load operating conditions. At 5\% load, LAPSA reduces total power loss by factors of 3$\times$ and 2.15$\times$ relative to, respectively, PWM and PFM operation, while preserving the nominal ripple envelope of 6\% inductor current ripple and 2\% output voltage ripple.

\section{Conclusion}\label{sec:conclusion}

A load-aware power system activation framework for DVPD architectures has been presented, enabling adaptive scaling of active VRs according to instantaneous load demand and spatial distribution. By selectively enabling regulators, the proposed approach mitigates excessive switching losses associated with conventional full-parallel PWM or PFM operation and sustains a high-efficiency plateau of approximately 87\% in the medium-to-light load regime. Practical implementation requires timely load sensing, supervisory computation, and coordinated gating across distributed regulators. Key hardware considerations include load sampling period, supervisory controller bandwidth, communication latency, and quantization resolution. %The 6~s latency observed in the SPICE--Python co-simulation reflects computational overhead of the external optimization loop and represents a conservative estimate of practical response time. 
In hardware realization, embedding the supervisory logic within FPGA- or microcontroller-based platforms is expected to substantially reduce response latency and enable real-time adaptation.

\bibliographystyle{IEEEtran}
\bibliography{./master_bib}

 \end{document}